\documentclass[12pt,thmsa]{article}
\usepackage{sw20lart}

\input tcilatex
\begin{document}

\author{Rolando Alvarado F$^a$., Maximo Ag\"{u}ero G$^b$.   \\
$^a$Universidad Autonoma de Zacatecas,\\
Centro de Estudios Multidisciplinarios, Apartado Postal 597-C, Zacatecas }
\title{An algebraic characterization of singular quasi--bi-Hamiltonian systems}
\maketitle

\begin{abstract}
In this paper we prove an algebraic criterion which characterizes singular
quasi-bi-hamiltonian structures constructed on the lines of a general,
simple, new formal procedure proposed by the authors. This procedure shows
that for the definition of a quasi-bi-hamiltonian system the requirement of
non-singular Poisson tensors, contained in the original definition by
Brouzet et al., is not essential. Besides, it is incidentally shown that one
method of constructing Poisson tensors available in the literature is a
particular case of ours. We present 2 examples.
\end{abstract}

\section{Introduction.}

The use of Hamiltonian methods in physics is as old as mathematical-physics
and can be traced back to the works of Euler and Lagrange in point and
continuum mechanics. Recently these methods, which we may divide in
symplectic and Poisson, have been developed in the works of Kostant\cite{kos}%
, Kirillov\cite{ki}, Lichnerowicz\cite{lili}, Guillemin\cite{gilo}, Soriau%
\cite{sop}, Weinstein\cite{we} and others in various fields which cover from
geometric quantization to control theory\cite{mardi}.

One current interest is now to construct Hamiltonian theories without the
help of a Lagrangian function\cite{hoji}, mainly because sometimes this way
of constructing the theory is not available (just remember the celebrated
Dirac theory of constraints). We mean, the hessian matrix of the Lagrangian
function has a rank less than the dimension of the configuration space,
hence, a Hamiltonian formulation does not seem available. Of course, the
equations of motion in the usual coordinates of the Lagragian are always, or
must be always, available. So, they can be considered as the starting point
in the construction of a Hamiltonian formulation. Clearly, if we have at
hand a given Hamiltonian formulation, it is interesting to know if we can
construct another one over this. However we must be clear as to what kind of
structure we wish to get on a manifold. It is possible to construct a
symplectic structure, which in the local coordinates of the symplectic
manifold give rise to the Lagrange brackets, or we may try to construct a
Poisson structure and, of course, the Poisson brackets in the local
coordinates of the Poisson manifold; which in the degenerate case cannot be
used to deduce, by inversion, the symplectic structure. In this paper we
will restrict ourselves to the discussion of a new and simply method to
construct singular quasi-bi-Hamiltonian structures; in the sense in which
Brouzet et al. defined this concept in their interesting paper\cite{bro};
with the help of an algebraic criterion deduced utilizing a decomposable
Poisson tensor. The main difference with the procedure of Brouzet et al. is
that we do not require the existence of any symplectic structure on the
manifold, we require only Poisson tensors which we allow to be singular
(degenerate); besides, our treatment is useful for any dimension of the
underlying manifold. Hence, the definition of a quasi-bi-Hamiltonian
structure is independent of any underlying symplectic structure. For this
reason we call the quasi-bi-Hamiltonian systems constructed ''singular''. We
will achieve this goal starting directly from a known Poisson tensor, in
order to erect another one on this basis. As we shall see, the method
provides a technique for defining the Hamiltonians for both structures, so
in our approach the only data that we need is a Poisson tensor. This is in
accordance with the usual procedures followed by some professional
constructors of Poisson structures\cite{hoji}\cite{god}, with just one
difference: some of them use a first order differential condition on the
Hamiltonian function (it is a constant of the motion) but we use a second
order condition, the Hamiltonian is the solution of a second order
differential equation, and we may consider that the usual first order
condition is a particular case of our condition.

In the next section (B) we give some useful definitions for the full
comprehension of the paper. In section (C) we give the main results,
contained in theorems (1) and (2).

In section (D) we give a brief introductory discussion of the Jacobi
structure from the point of view of our general methodology, in section (E)
we give 3 examples of the method and in the last section (F) we give the
conclusions, which try to iluminate the full disscusion in the article.

\section{The definition of Poisson structures.}

Le $M\;$be a smooth manifold and $C^\infty (M,\Re )$ the ring of all real
valued, infinitely differentiable, functions on $M\;$\cite{mardi}\ . A
Poisson structure on $M$ is given by a bilinear operation: $\{,\}$ on $%
C^\infty (M,\Re )$ such that the maps: $X_H=\{H,*\}$, $L_H=\{*,H\}$ are
derivations, this operation is known as the '' Poisson bracket''. A manifold
endowed with a Poisson bracket on $C^\infty $ is called a Poisson manifold
and we will denote it by the pair: $O=<M,\{,\}>$.

We may understand by a Poisson structure on a manifold the explicit
definition of a Poisson bracket, or the definition of a contravariant
antisymmetric 2-tensor defined at all the points of $M$. In local
cosymplectic coordinates this tensor is:

\begin{equation}
\left\{ F,G\right\} =J^{ij}(x)\frac{\partial F}{\partial x^i}\frac{\partial G%
}{\partial x^j}.  \tag{1}
\end{equation}
Usually, to specify the 2-tensor we use the fundamental Poisson brackets of
the local coordinates: $J^{ij}(x)=\{x^i,x^j\}$. The Jacobi identity gives us
a first order partial differential equation which the 2-tensor must satisfy,
but we will not display it here.

The tensor defines an isomorphism between the cotangent bundle $T^{*}M$ and
the tangent bundle $TM$ if, and only if, it is non-degenerate. However, we
will not consider this as an essential condition in the definition of the
Poisson structure, because, as we will see, it is not necessary for the
construction of the quasi-bi-Hamiltonian structure. If we use the 2-tensor,
we will use the pair: $O=$ $<M,J>$ to express in a coordinate-free manner
the Poisson manifold.

DEFINITION 1.- We will say that we have given a Hamiltonian structure on the
Poisson manifold $O$ when we give the scalar generator of the dynamics $H\in
C^\infty (M,\Re )$, which is known as the Hamiltonian of the Hamiltonian
structure. We will denote this by the triplet: $K_H=$ $<O,H>$ $=$ $%
<M,\{,\},H>$ .

To define the dynamics in the local coordinates of $O$ we need only the
Poisson bracket and the Hamiltonian, because with these elements we may
define a dynamics, locally or globally, as the integral curves of the
Hamiltonian vector field given by:

\[
X_H=\{x^i,x^j\}\frac{\partial H}{\partial x^i}\frac \partial {\partial x^j} 
\]
in local coordinates. So, in all the constructions of a Poisson structure it
is important to define the set of all possible Hamiltonians.

DEFINITION 2.-\cite{bue}\ We will say that we have a bi-Hamiltonian
structure if, and only if, given a Hamiltonian structure: $K_1=$ $%
<M,\{,\}_1,H_1>$, it is possible to construct another different Hamiltonian
structure: $K_2=$ $<M,\{,\}_2,H_2>$ such that: $\{x^i,H_1\}_1=\{x^i,H_2\}_2$%
. So, the dynamics accept two different formulations:

\begin{equation}
\frac{dx^i}{dt}=\left\{ x^i,H_1\right\} _1=\left\{ x^i,H_2\right\} _2\text{.}
\tag{2}
\end{equation}
The attempt of construction of this kind of formulations, even in the
singular (degenerate) case, have led to the development of a series of
techniques for its effective realization\cite{bue}.

The previous definitions are standard in the theory of bi-Hamiltonian
systems; now we must pass to the quasi-bi-Hamiltonian case of Brouzet et al.%
\cite{bro} but, as we wish to treat the singular case, we give a modified
definition of a quasi-bi-Hamiltonian system in such a way that the
singularities are not important:

DEFINITION 3.- Let $K_H$ be a Hamiltonian system. We say that it admits a
quasi-bi-Hamiltonian structure if, and only if:

( i ).- There exists on $M$ a Poisson tensor $J_{\wedge }$ compatible with $%
J $, i. e. its Schouten bracket (for the definition and properties of the
Schouten bracket we use the book by Marsden and Ratiu\cite{mardi}) commutes.

( ii ).- There exists a non-vanishing function $\rho \in C^\infty (M,\Re )$
such that: $\rho $ $\left( dH\rfloor J\right) $ is a globally Hamiltonian
vector field.

Here $\rfloor $ denotes the contraction operation on a tensor field. We
shall denote the quasi-bi-Hamiltonian system with:

$QBH=$ $<K_H,J_{\wedge },\rho >$ $=$ $<M,$ $J,$ $H,$ $J_{\wedge },$ $\rho >$%
. We will see later how to get the second Hamiltonian.

If $\rho \left( dH\rfloor J\right) $ is globally Hamiltonian\cite{jost} for $%
J$, i. e. $(\exists F):\rho \left( dH\rfloor J\right) =X_F$, then $QBH$ is
called ''exact''. As we shall see later, the function $F$ is the solution of
a partial differential equation of first order, and we shall understand the
solutions to this equation in a local sense, as given by the usual theorems
of existence\cite{hil}. In this way we can make the following difference: if
the solution is available in closed form the vector field $\rho \left(
dH\rfloor J\right) $ is globally Hamiltonian, but if the solutions are only
available through the theorems of existence, usually power series which
converge in some open disk, we shall call the vector field $\rho \left(
dH\rfloor J\right) $ a locally Hamiltonian vector field\footnote{%
In the symplectic framework we call a Hamiltonian vector field
\par
a locally Hamiltonian vector field if, and only if, its contraction with the
symplectic two-co-tensor is zero. In symbols, if $\Omega $ is the
two-co-tensor and $X$ the vector field we have: $d($ $X\rfloor \Omega )=0$
and by Poincare lemma we have that locally: $X\rfloor \Omega =dH$[see ref%
\cite{mardi}. p.141]. Hence our introduced notion for the vector field $\rho
(dH\rfloor J)$ is not arbitrary.}.

The definitions in terms of symplectic structures, hence with the recursion
operator at hand (Nijenhuis operator), were given by R. Brouzet et al.\cite
{bro} and can be deduced from (i) and (ii) if both Poisson tensors are
non-degenerate, because in such a case we have: $\omega =-J^{-1}$, $\omega
_{\wedge }=-J_{\wedge }^{-1}$ hence the Nijenhuis operator is: $-J_{\wedge
}\circ \omega $. Definition (3) can be changed by that of Brouzet et al.
just replacing the Schouten bracket with the Nijenhuis torsion for the test
of the compatibility of the symplectic structures (which is, of course, more
complicated from the point of view of calculations than the relatively
easier Schouten bracket). Hence, in this sense the extension of our
definition covers that one of Brouzet et al., because even when the
symplectic structure is not available, our definition has no problem. It is
necessary to remark that the definition is introduced because it shows that
the necessity of an inverse for the Poisson structure is not an important
condition.

\section{Compatibility of Poisson structures.}

The notation in this section is as follows: $\lceil ,\rceil $ denotes the
Schouten bracket. We start with the well-known:

DEFINITION 4.-\cite{cari} A Poisson tensor, $\wp $, will be called
decomposable (or the Poisson tensor $J_1$ extendable if there is a Poisson
tensor $J_2$ such that we may compose the tensor $\wp $ ) if, and only if,
it is possible to write it as: $\wp =J_1+J_2$ where each tensor $J_1$,$\;J_2$
is a Poisson tensor and: $\lceil J_1,J_2\rceil =0$.

We will call the tensors which commute under the Schouten bracket:
compatible tensors. As is well known\cite{cari}, extensions of Poisson
tensors can be used to study the problem of classification of solvable Lie
algebras.

The methodology which we will use here to extend Poisson tensors is as
follows: given a Poisson tensor of the form $J=X_1\wedge X_2$, with the two
vector fields: $X_1=f_i\frac \partial {\partial x^i}$, $X_2=g_i\frac
\partial {\partial x^i}$ and a third vector field $X_3$, we construct the
new Poisson tensor $J_{\wedge }$ as $X_\gamma \wedge X_3$, where $X_\gamma
=\gamma $ $\rfloor $ $J$ and $\gamma $ is a 1-co-tensor which we shall
consider of the integrable (exact) form: $\gamma =dH$, with $H$ a real
valued function. Note that the choice of Poisson tensors is not general,
however, this lack of generality is presented in many constructions of this
kind.

So, if we follow the methodology just described to construct a new Poisson
tensor over an old one, we only need to show that the new Poisson tensor
commutes with the old one, and this procedure will give us the conditions to
determine the vector field $X_3$ (the conditions obtained are independent of
the representation used by the generators of the algebra.). Before proceed
to the explicit constructions it is very important to see that in the
methodology one crucial step is the choice of the form of two objects:

\textbf{(A)}.-the Poisson tensor $J=X_1\wedge X_2$ which is not the most
general one, but however is the form used by Hojman (see ref.\cite{hoji}
p.669. In the next section we establish the explicit relation with the
method by Hojman. Even in the case of J. Goedert\cite{god} are requiered
some restrictions, specifically the space dimension).

\textbf{(B)}.- The choice of the vector field $X_H=dH$ $\rfloor $ $J$, as
the contraction of the Poisson tensor.

These suppositions are independent in the sense that it is only in the
lemma4 and the identity (7) that both are used to get a reduced set of
conditions involving the vectors $X_1$, $X_2$, $X_3$, in such a way that it
is possible to know the third vector in terms of the first two, i. e. the
second Poisson tensor with just the elements of the first one. We remark
this because, as we shall see in the next section, with this choice we can
re-construct the quasi-bi-Hamiltonian system in a very easy way.

We start with a few easy lemmas (in the development we shall suppose that
the vector fields $X_1$, $X_2$, $X_3$ are linearly independent. There are no
conditions on $X_H$ unless otherwise stated):

LEMMA1.- The 2-contra-tensors $X_1\wedge X_2$, $X_H\wedge X_3$ are Poisson
tensors, respectively, if, and only if:

\begin{eqnarray}
\left[ X_1,\;X_2\right] &=&N_1X_1+N_2X_2  \tag{3a} \\
\left[ X_H,\;X_3\right] &=&A_1X_H+A_2X_3,  \tag{3b}
\end{eqnarray}
where $N_1$, $N_2$, $A_1$, $A_2$ are arbitrary functions of the coordinates.

PROOF: the Schouten brackets for each tensor are:

\begin{eqnarray*}
&&X_1\wedge \left[ X_1,\text{ }X_2\right] \wedge X_2=0 \\
&&X_H\wedge \left[ X_H,\text{ }X_3\right] \wedge X_3=0
\end{eqnarray*}
So, we can see that the conditions (3a-b) are sufficient conditions, because
if they hold, then the Schouten bracket vanishes, as a simple calculation
shows. To the opposite side: if the Schouten bracket is zero, then the
conditions (3a-b) are the only solutions.$\bullet $

LEMMA2.- $X_H$ is an infinitesimal automorphism of the Poisson tensor $%
X_1\wedge X_2$ if, and only if ( to avoid any confusion: we use the notation 
$X_{H\text{ }}$for convenience, it does not mean that the vector field is
Hamiltonian):

\begin{eqnarray}
\left[ X_H,\;X_1\right] &=&-C_1X_1+B_2X_2  \tag{4a} \\
\left[ X_H,\;X_2\right] &=&C_1X_1+C_2X_2  \tag{4b}
\end{eqnarray}
where $C_1$, $C_2$, $B_2$ are arbitrary functions of the coordinates.

PROOF:

The Lie derivative of the tensor $X_1\wedge X_2$ with respect to the vector
field $X_H$ is, in terms of the Schouten bracket:

\[
\lceil X_H,\text{ }X_1\wedge X_2\rceil =\left[ X_H,\text{ }X_1\right] \wedge
X_2+X_1\wedge [X_H,\text{ }X_2]=0 
\]
clearly if $X_H$ is an infinitesimal automorphism this Lie derivative must
be zero. We can see that the conditions (4a-b) are sufficient, because if
they hold then the Lie derivative is zero, as a substitution shows. They are
necessary too because if the Schouten bracket vanishes, the only solution
for it is through conditions (4a-b).$\bullet $

LEMMA3.-Suppose that $X_H$ is an infinitesimal automorphism of the tensor $%
X_1\wedge X_2$, then this tensor and the tensor $X_H\wedge X_3$ are
compatible tensors if, and only if:

\begin{eqnarray}
\left[ X_3,\;X_1\right] &=&D_1X_H+D_2X_2  \tag{5a} \\
\left[ X_3,\;X_2\right] &=&E_1X_H+E_2X_2  \tag{5b}
\end{eqnarray}
PROOF: The Schouten bracket of the tensors is:

\[
\left( \left[ X_H,\;X_1\right] \wedge X_2+X_1\wedge \left[ X_H,\;X_2\right]
\right) \wedge X_3-X_H\wedge \left( \left[ X_3,\;X_1\right] \wedge
X_2+X_1\wedge \left[ X_3,\;X_2\right] \right) , 
\]
but by the lemma2 (which we suppose to hold, i. e. the relations 4a-b are
valid) the first term is zero, hence we get the equation:

\[
X_H\wedge \left( \left[ X_3,\text{ }X_1\right] \wedge X_2+X_1\wedge \left[
X_3,\text{ }X_2\right] \right) =0. 
\]
So, the result (5a-b) is an immediate consequence, because if the conditions
(5a-b) hold, the Schouten bracket vanishes, and if the Schouten bracket
vanishes the only solution to the equation is through conditions (5a-b).$%
\bullet $

THEOREM1.- Given four vectors $X_1$, $X_2$, $X_3$, $X_H$, then the tensors $%
X_1\wedge X_2$, $X_H\wedge X_3$ are compatible Poisson tensors such that $%
X_H $ is an infinitesimal automorphism of $X_1\wedge X_2$ if, and only if,
they form the basis of an algebra with the following commutation relations:

\begin{eqnarray}
\left[ X_1,\text{ }X_2\right] &=&N_1X_1+N_2X_2  \tag{6a} \\
\left[ X_H,\text{ }X_3\right] &=&A_1X_H+A_2X_3  \tag{6b} \\
\left[ X_H,\text{ }X_1\right] &=&-C_2X_1+B_2X_2  \tag{6c} \\
\left[ X_H,\text{ }X_2\right] &=&C_1X_1+C_2X_2  \tag{6d} \\
\left[ X_3,\text{ }X_1\right] &=&D_1X_H+D_2X_2  \tag{6e} \\
\left[ X_3,\text{ }X_2\right] &=&E_1X_H+E_2X_1  \tag{6f}
\end{eqnarray}
PROOF.- just use the lemmas (1-2-3).$\bullet $

In this way the notions of compatibility of Poisson tensors and the property
of one vector of being an infinitesimal automorphism of one Poisson tensor
are algebraic concepts whose structure is contained in the relations (6a-6f).

This is a 4-dimensional algebra which we shall reduce to a 3-dimensional one
with the help of the condition: $X_H=X_1\left( dH\right) X_2-X_2\left(
dH\right) X_1$. We must remark that this condition has not been used in the
development at any point, so it is an independent condition. Next, we shall
use the identity:

\begin{equation}
\left[ X_H,\;X_i\right] =X_1(dH)\left[ X_2,\text{ }X_i\right] +X_2\left(
dH\right) \left[ X_1,\text{ }X_i\right] -X_iX_1\left( dH\right)
X_2-X_iX_2\left( dH\right) X_1  \tag{7}
\end{equation}
which follows from the derivation property of the commutator and the form
which we have supposed for the vector field $X_H$. Note that the Lie bracket
is not $C^\infty (M,\Re )$-bilinear and that the identity (7) change if we
use another form for the infinitesimal automorphism.

LEMMA4.- If $X_H=X_1(dH)X_2-X_2(dH)X_1$ the 4-dimensional algebra (6a-f) is
reduced to the 3-dimensional algebra (of 3 linearly independent vectors):

\begin{eqnarray*}
\left[ X_1,\text{ }X_2\right] &=&N_1X_1+N_2X_2 \\
\left[ X_3,\text{ }X_1\right] &=&\left( D_1X_1(dH)+D_2\right)
X_2-D_1X_2(dH)X_1 \\
\left[ X_3,\text{ }X_2\right] &=&\left( E_2-E_1X_2(dH)\right)
X_1-E_1X_1(dH)X_2
\end{eqnarray*}
if we choose the functions:

\begin{eqnarray*}
C_1 &=&X_2(dH)N_1-X_2^2(dH) \\
C_2 &=&X_1X_2(dH)-X_1(dH)N_1 \\
B_1 &=&-C_2 \\
B_2 &=&\frac{X_1(dH)}{X_2(dH)}\left( X_1X_2(dH)-X_1(dH)N_1\right) +\frac{%
X_2X_1(dH)}{X_2(dH)}-X_1^2(dH) \\
N_2 &=&\frac{X_1X_2(dH)}{X_2(dH)}-\frac{X_1(dH)}{X_2(dH)}N_1+\frac{X_2X_1(dH)%
}{X_2(dH)} \\
A_1 &=&\frac{X_1(dH)}{X_2(dH)}E_2-X_2(dH)D_1-X_1(dH)E_1+\frac{X_3X_2(dH)}{%
X_2(dH)} \\
A_2 &=&-\left( X_2(dH)D_2+\frac{\left( X_1(dH)\right) ^2}{X_2\left(
dH\right) }E_2+X_3X_1(dH)+\frac{X_3X_2(dH)X_1(dH)}{X_2\left( dH\right) }%
\right)
\end{eqnarray*}
leaving undetermined the remaining functions: $D_1$, $D_2$, $E_1$, $E_2$, $%
N_1$.

PROOF.-The idea used here is quite simple: we shall try to satisfy
identically the commutation relations (6b-c-d) with the help of the form
which we use for the vector field $X_H$ and an adequate selection of the
arbitrary functions.

So, with this in mind, we use the identity (7) and the form of $X_H$ to
change the algebra (6a-f) to: 
\begin{equation}
\left[ X_1,\text{ }X_2\right] =N_1X_1+N_2X_2  \tag{8a}
\end{equation}

\begin{equation}
\begin{tabular}{lll}
$X_1(dH)\left[ X_2,\text{ }X_3\right] -X_3X_1\left( dH\right) X_2+X_2\left(
dH\right) \left[ X_1,\text{ }X_3\right] -X_3X_2(dH)X_1=$ &  &  \\ 
$A_2X_2+A_1(X_1(dH)X_2-X_2(dH)X_1)$ &  & 
\end{tabular}
\tag{8b}
\end{equation}

\begin{equation}
X_1(dH)\left[ X_2,\text{ }X_1\right]
-X_1^2(dH)X_2-X_1X_2(dH)X_1=-C_2X_1+B_2X_2  \tag{8c}
\end{equation}
\begin{equation}
-X_2X_1(dH)X_2+X_2(dH)\left[ X_1,\text{ }X_2\right]
-X_2^2(dH)X_1=C_1X_1+C_2X_2  \tag{8d}
\end{equation}

\begin{equation}
\left[ X_3,\text{ }X_1\right] =D_1\left( X_1\left( dH\right) X_2-X_2\left(
dH\right) X_1\right) +D_2X_2  \tag{8e}
\end{equation}

\begin{equation}
\left[ X_3,\text{ }X_2\right] =E_1\left( X_1(dH)X_2-X_2(dH)X_1\right) +E_2X_1
\tag{8f}
\end{equation}

Now we put (8a) in (8c) and (8d) and (8e), (8f) in (8b). From the
substitution of (8a) in (8c) and (8d) we get the equations:

\begin{eqnarray*}
(X_1(dH)N_1+C_2-X_1X_2(dH))X_1+(X_1(dH)N_2-X_1^2(dH)-B_2)X_2 &=&0 \\
(X_2(dH)N_1-X_2^2(dH)-C_1)X_1+(X_2(dH)N_2-C_2-X_2X_1(dH))X_2 &=&0
\end{eqnarray*}
so, by the linear independence of the vector fields at each point of the
manifold, we get the equations:

\begin{eqnarray}
X_1(dH)N_1+C_2-X_1X_2(dH) &=&0  \tag{9a} \\
X_1(dH)N_2-B_2-X_1^2(dH) &=&0  \tag{9b} \\
X_2(dH)N_1-C_1-X_2^2(dH) &=&0  \tag{9c} \\
X_2(dH)N_2-C_2-X_2X_1(dH) &=&0  \tag{9d}
\end{eqnarray}
Now, we put (8e), (8f) in (8b) to get:

\[
\begin{tabular}{ll}
$A_1\left( X_1(dH)X_2-X_2(dH)X_1\right) +A_2X_2=$ &  \\ 
$-X_1\left( dH\right) \left( E_1\left( X_1(dH)X_2-X_2(dH)X_1\right)
+E_2X_1\right) -X_3X_1\left( dH\right) X\;-$ &  \\ 
$X_2\left( dH\right) \left( D_1\left( X_1\;\left( dH\right) X_2-X_2\left(
dH\right) X_1\right) +D_2X_2\right) -X_3X_2(dH)X_1$ & 
\end{tabular}
\]
grouping the terms in this equation we get the expression:

\begin{eqnarray*}
&&(-(X_1\;\left( dH\right) )^2E_1-X_3X_1\left( dH\right) -X_2\left(
dH\right) D_1X_1\left( dH\right) - \\
&&X_2\left( dH\right) D_2-A_1X_1\left( dH\right) -A_2)X_2+ \\
&&(X_1\left( dH\right) E_1X_2\left( dH\right) -X_1\left( dH\right)
E_2+(X_2\;\left( dH\right) )^2D_1- \\
&&X_3X_2(dH)+A_1X_2(dH))X_1
\end{eqnarray*}
which is zero. Again, the linear independence of the vector fields give us
two equations:

\begin{equation}
\begin{tabular}{lll}
$A_2+A_1X_1(dH)+D_2X_2(dH)+D_1X_1(dH)X_2(dH)+E_1\left( X_1(dH)\right) ^2+$ & 
&  \\ 
$X_3X_1(dH)=0$ &  & 
\end{tabular}
\tag{10a}
\end{equation}

\begin{equation}
\begin{tabular}{lll}
$A_1X_2(dH)+D_1\left( X_2(dH)\right) ^2-E_2X_1\left( dH\right) +E_1X_1\left(
dH\right) X_2\left( dH\right) -$ &  &  \\ 
$-X_3X_2\left( dH\right) =0$ &  & 
\end{tabular}
\tag{10b}
\end{equation}
The equations (9a-d) and (10a-b) are what we need. From (9a) and (9c) we get 
$C_1$, $C_2$. Using these two functions we get, with the help of (9b) and
(9d) the functions $B_1$, $N_2$ given in the lemma. From (10b) we get the
function $A_1$. We put this function in (10a) to get $A_2$. Hence the lemma.$%
\bullet $

The algebra given in the lemma 4 is such that its representations allow us
to construct extensions for Poisson tensors. However, it is complicated, but
fortunately, we can choose the functions $N_1$, $E_1$, $E_2$, $D_1$, $D_2$,
in such a way that the algebra becomes easier to treat.

THEOREM2.-Three linearly independent vector fields $X_1$, $X_2$, $X_3$, and
a fourth vector $X_H$ constructed as the contraction with an exact
1-co-tensor of the 2-tensor $X_1\wedge X_2$, allow us to construct two
compatible Poisson tensors if they are the base of the algebra with
commutation rules:

\begin{eqnarray*}
\lbrack X_1,\text{ }X_2] &=&0 \\
\lbrack X_3,\text{ }X_1] &=&X_1-X_2 \\
\lbrack X_3,\text{ }X_2] &=&0
\end{eqnarray*}
besides, the hamiltonian $H$ of the Poisson structure $X_1\wedge X_2$
satisfies the second order, factorizable, differential equation: $%
X_1X_2(dH)=0$.

PROOF.- The idea here is, again, quite simple: the lemma4 give us a
3-dimensional algebra obtained with two hypothesis: one vector is of a form
such that only three vectors are requiered, and we choose the coefficients
in such a way that three commutation rules are identities. Hence, now we
shall choose the remaining coefficients and we shall use a new condition.

The choice for the coefficients is:

\begin{eqnarray*}
N_1 &=&E_1=E_2=0, \\
D_1 &=&-\frac 1{X_2\left( dH\right) },\text{ }D_2=-1+\frac{X_1\left(
dH\right) }{X_2\left( dH\right) }
\end{eqnarray*}
and the new condition is:

\[
N_2=0=\left( X_1X_2\;+\;X_2X_1\right) \left( dH\right) 
\]
From this condition, and because $[X_1,$ $X_2]=0$, for the Hamiltonian we
get the equation : $2X_1X_2\left( dH\right) =0$. Choosing the coefficients
in the way which we have indicated reduce the algebra given in the lemma4 to
the algebra given in the theorem 2. Hence, the theorem 2 is proved.$\bullet $

We must remark that the other functions $A_1,A_2,B_1,B_2,C_1,C_2$ do not
give us any condition, because we have not imposed any on them. Well, in
fact there are many ways in which we can choose the functions, as must be
clear, the one which we offer here is used because it is very easy to
construct the representations in terms of first order differential
operators, as we shall see in the example. As many possibilities are open,
the notation for the algebra of the lemma 4 could be: $\Im \left(
N_1,E_1,E_2,D_1,D_2\right) $, so, the case which we shall treat is:\ $\Im
\left( 0,0,0,-\frac 1{X_2\left( dH\right) },-1+\frac{X_1\left( dH\right) }{%
X_2\left( dH\right) }\right) =_{def}\Delta $. The treament which we have
given is useful for Poisson tensors of the monomial form $X_1\wedge X_2$,
but our aim is not a general method to extend arbitrary Poisson tensors,
instead, we are trying to construct singular-quasi-bi-Hamiltonian systems,
as we shall make in the next section. It is very important to remark that
our algebraic criterion is sufficient, for this reason the consideration of
the necesary conditions is not so important.

\section{The quasi-bi-hamiltonian system}

With our decomposable Poisson tensor, $\wp =X_1\wedge X_2+X_H\wedge X_3$,
available through an algebraic criterion, is the moment of constructing the
quasi-bi-Hamiltonian system. For this end we will use the infinitesimal
automorphisms of the Poisson structures, which are given by:

\begin{eqnarray}
X_H &=&dH\text{ }\rfloor \text{ }\left( X_1\wedge X_2\right) =X_1\left(
dH\right) X_2-X_2\left( dH\right) X_1,  \tag{11a} \\
X_F &=&dF\text{ }\rfloor \text{ }\left( X_H\wedge X_3\right) =X_H\left(
dF\right) X_3-X_3\left( dF\right) X_H.  \tag{11b}
\end{eqnarray}
Hence a relation between the different Hamiltonian vector fields is:

\begin{equation}
X_F=\left\{ H,F\right\} X_3+\rho (F)X_H,\text{ }\rho (F)=-X_3(dF)\text{.} 
\tag{12}
\end{equation}
So, if we choose $F$ as an integral, or even as a Casimir, of the first
Poisson tensor (Brouzet et al. only used the integrals, because the Casimir
functions are not allowed for them), we easily get: 
\begin{eqnarray}
X_F &=&\rho X_H=\rho \left( \text{ }dH\text{ }\rfloor \text{ }J\text{ }%
\right) ,  \tag{13a} \\
X_F\left( dF\right) &=&0  \tag{13b} \\
X_F\left( dx^i\right) &=&\rho X_H\left( dx^i\right)  \tag{13c}
\end{eqnarray}
As required by the definition (3). This procedure is, of course, suggested
by Brouzet et al.\cite{bro}[p. 2070-2071 eq. (4)]. Hence we have constructed
an exact quasi-bi-Hamiltonian system on the basis of a decomposable Poisson
tensor. For our quasi-bi-Hamiltonian dynamical system the property of being
Pfaffian is not available, because this is defined by Brouzet et al. in
connection with the Nijenhuis operator\cite{bro}\ . Now, we must remark here
that the relations (11a-b) and the reduction to the form (13a) are the main
motivations for the use of the suppositions (A) and (B) of the former
section.

In this way we have, in general, constructed a quasi-bi-hamiltonian system
on the basis of the algebra $\Im (N_1,E_1,E_2,D_1,D_2)$. We can see that,
for example, the determination of the first Hamiltonian in terms of the
available elements, the vector fields, is possible in the case of the
algebra $\Delta $, and of course, if we have the form of this Hamiltonian,
the second Hamiltonian is in principle, known.

Hence we have the family of $QBH$:

\begin{equation}
QBH_p=\left\langle M,\text{ }X_1\wedge X_2,\text{ }H,\text{ }X_H\wedge X_3,%
\text{ }F\right\rangle  \tag{14}
\end{equation}
where in the last coordinate, instead of the function $\rho $, we use the
second Hamiltonian because it is now available. There is a possible
confusion here as $F$ is, in fact, local, due to its definition as an
integral of the first Poisson tensor (we mean: it is the solution of a
partial differential equation of first order, and we understand this
solution in the local sense). However, it is possible to find it explicitly
in closed form in some simple cases. We will, nevertheless, restrict
ourselves to this case. The algebra $\Delta $ is clearly solvable.

We can note an interesting fact of the algebra $\Delta $: if $X_2=0$, then
the differential equation: $\dot{x}_i=X_1(dx_i)$ by Lie theorem is
integrable by cuadratures\cite{Arni}, if it is $2$-dimensional. Now, let us
establish the connection with the Hojman method to construct Hamiltonian
theories for autonomous first-order differential systems\cite{hoji}.

For the $n$-dimensional case consider the equation: $\dot{x}_i=X_1\left(
dx_i\right) .$This differential equation admits a Poisson structure of the
form $J=$ $X_1\wedge X_3$, by just choosing the Hamiltonian as $X_1(dH)=0$
(clearly a particular case of the differential equation: $X_1X_2\left(
dH\right) =X_2X_1\left( dH\right) =0$ and $X_1(dH)=cte.$) because by
contraction we get the vector field: $X_3(dH)X_1$. Under this condition $%
\rho =X_3(dH)$ is a constant of the motion (PROOF$:X_3\left( X_1(dH)\right)
-X_1\left( X_3(dH)\right) =X_1\left( dH\right) \Rightarrow X_1\left(
X_3\;\left( dH\right) \right) =0$ )\ as required by Hojman method. Hence the
scaling used by Hojman ( see ref.\cite{hoji} p. 669 equation (12) ) for the
vector field $\bar{X}_1=\rho X_1$ (in components of the vector fields it is: 
$\bar{\eta}^i=\rho \eta ^i$ ) is possible. For this reason the Hamiltonian
structure is for the system with vector field of the form: $\bar{X}%
_1=X_1\rho $ which can be seen as Hamiltonian. The Hamiltonian vector field
constructed by Hojman is, thus: $dH$ $\rfloor $ $J$ $=\bar{X}_H=\rho X_1$.
For this reason, and because $X_2=0$ can be seen as a particular case of the
algebra $\Delta $. Then, our construction covers that one of Hojman.

This is the unique way in which the connection can be established, because
it is possible to commit the mistake of trying to apply the procedure which
we offer to construct an extended Poisson tensor over that constructed by
the method of Hojman. If we do this we get a contradiction. Let us show
this: Let $X_1\wedge X_3\;$be the Poisson tensor obtained by the Hojman
method, then, our procedure gives us the extension with the help of the
tensor: $\bar{X}_H\wedge Y$, and the conditions:

\begin{eqnarray*}
\left[ Y,X_1\right] &=&X_1-X_3, \\
\left[ X_1,X_3\right] &=&0, \\
\left[ X_3,Y\right] &=&0.
\end{eqnarray*}
But, according to the Hojman method we must have: $\left[ X_3,X_1\right]
=X_1 $, which cannot be satisfied if we use the methodology proposed in the
method which we offer. Hence, this procedure is wrong.

So, we reject any criticism to our method on the basis of the preceding
calculation, which comes from an incorrect consideration. The only way to
establish the connection is that which considers the Hojman method as a
particular case of ours in the lines which we have already explained. Let us
make a last comment: Hojman proposed in his paper\cite{hoji} the idea that
if the differential equation in question has, say, $m$-symmetries ($m\leq n$%
), then, on the basis of his method, entirely based on the assumption $X_2=0$
for the algebra $\Delta $, we can get new Poisson tensors. So, we must find
a representation for the algebra:

\begin{equation}
\left[ X_i,X_1\right] =X_1,i=1,...,m.  \tag{15}
\end{equation}
Hence, if $m=n$ we get the result (not remarked in Hojman paper) that any $n$%
-dimensional integrable system: $\dot{x}_i=X_1(dx_i)$, admits $n$ Poisson
structures. The vector fields $X_i$ are of course, the classical Lie
symmetries of the differential equation. In fact, Hojman proposed (ref \cite
{hoji} p. 673) a way to extend the Poisson tensor constructed by his method,
but based on the relation (15) and, because this relation is deduced from
our algebra $\Delta $, the Poisson tensor extended by Hojman method are
compatible.

If we want to get a singular bi-Hamiltonian system for the 2-dimensional
case (1-dimensional if we use Darboux coordinates) we must add the
condition: $X_3(dF)=-1$, to the condition $\{H,F\}=0$, to get the second
hamiltonian $F$, which shows that this case is more restricted, from the
point of view of our calculations, than the singular one.

It is important to remark two points:

(1).- The method which we propose to construct decomposable Poisson tensors
is, essentially, a problem of algebra representations\cite{cari}, because
any realization of the Lie algebra $\Delta $ can be used to construct a
decomposable Poisson tensor. The infinitesimal authomorphisms which leave
invariant our decomposable Poisson tensor $\wp $ are generated, as before,
by the vector fields: $X_\gamma =\gamma $ $\rfloor $ $\wp $ with $\gamma $
any 1-form.

(2).-If we choose such vectors so that the commutation relations which
define $\Im $ are not realized, our Poisson tensors are not compatible and
thus, we do not have the possibility of reducing the full algebra to $\Delta 
$ which gives us the operative formulation for the construction of a
quasi-bi-Hamiltonian dynamical system.

\section{Digression on Jacobi structures.}

It is not our aim to discuss in full detail the Jacobi structures\cite{ki}%
\cite{lili}, but we think that it is important to remark how the methodology
which we propose for the extension of Poisson tensors could be useful for
constructing Jacobi structures.

We start with a general definition ( see ref. \cite{chino} p. 6314 ):

DEFINITION 5.- On a smooth ( $C^\infty $) manifold $M$ we say that we have a
Jacobi structure if, and only if, it is possible to construct a pair $%
<\Lambda ,$ $X_H>$ such that $\Lambda $ is a 2-contra-tensor and $X_H$ is
1-contra-tensor on $M$ such that the following conditions holds:

\begin{equation}
\lceil \Lambda ,\text{ }\Lambda \rceil =2X_H\wedge \Lambda \text{, }\lceil
X_H\text{, }\Lambda \rceil =0  \tag{16}
\end{equation}
We use the notation $X_H$ to connect with the former sections. Clearly, this
is not a Hamiltonian vector field unless otherwise stated, it is an
arbitrary vector field. Our methodology is summarized in the following:

THEOREM\ 3.- Three linearly independent vector fields $X_1$, $X_2$, $X_H$,
allows us to construct a Jacobi structure of the form $<X_1\wedge X_2,$ $%
X_H> $ on a smooth manifold $M$ if they satisfy the commutation rules:

\begin{eqnarray}
\left[ X_1,\text{ }X_2\right] &=&-X_H  \tag{16a} \\
\left[ X_H,\text{ }X_1\right] &=&-AX_1+BX_2  \tag{16b} \\
\left[ X_H,\text{ }X_2\right] &=&CX_1+AX_2  \tag{16c}
\end{eqnarray}
with $A$, $B$, $C$, arbitrary functions.

PROOF: The assertion in the theorem just requires the sufficiency of the
commutation rules (16a-c) to be shown, that is: if the vector fields form
the algebra with commutation rules (16a-c), which we shall denote as $\jmath
(A,B,C)$, then we can construct the pair $\left\langle X_1\wedge
X_2,X_H\right\rangle $. The proof is as follows: expand the Schouten bracket
to get the equation:

\[
2X_1\wedge [X_1,\text{ }X_2]\wedge X_2=2X_H\wedge (X_1\wedge X_2) 
\]
so, according to the definition we must have:

\[
\left[ X_1,X_2\right] =-X_H 
\]
the other condition simply states that the vector field $X_H$ is an
infinitesimal automorphism of the 2-contra-tensor. We meet sufficient
conditions for this requierement in lemma2, the conditions are equal to
(16a-b), but we changed the arbitrary functions to avoid confusion. Hence,
if these algebraic conditions are satisfied, we can construct a Jacobi
structure on the manifold, by the conditions in the definition. The theorem
is proved.$\bullet $

We shall give 1 example below.

\section{Examples.}

The key points for constructing an example of an exact quasi-bi-Hamiltonian
system are: a Poisson tensor, its contraction with an integrable 1-form and
an arbitrary vector. The main source of all the examples is the work of
Cari\~{n}ena et al.\cite{cari} on solvable Lie algebras.

We will consider three ortogonal vectors, for the sake of generality and
non-triviality. In this case the algebra $\Delta $ to be satisfied is:

\begin{eqnarray*}
\lbrack X_1,\text{ }X_2] &=&0 \\
\lbrack X_3,\text{ }X_1] &=&X_1-X_2 \\
\lbrack X_3,\text{ }X_2] &=&0
\end{eqnarray*}
In each example we will use a super-index to denote the particular
realization of the algebra and the number of the example.

1.- Every 2-dimensional example is trivial, because the Schouten brackets
are 3-tensors which vanish in two dimensions, by construction.

2.-Consider a 3-dimensional (the first non-trivial dimension for a 3-vector
like the Schouten bracket) example with the Poisson tensor:

\begin{eqnarray}
J^{(2)} &=&(x_1\frac \partial {\partial x_2}-x_2\frac \partial {\partial
x_1})\wedge \frac \partial {\partial x_3},  \tag{21a} \\
X_1^{(2)} &=&x_1\frac \partial {\partial x_2}-x_2\frac \partial {\partial
x_1},X_2^{(2)}=\frac \partial {\partial x_3}  \tag{21b}
\end{eqnarray}
and the arbitrary vector:

\begin{equation}
X_3^{(2)}=P_1(x_1,x_2)\frac \partial {\partial x_1}+P_2(x_1,x_2)\frac
\partial {\partial x_2}+P_3(x_1,x_2)\frac \partial {\partial x_3}.  \tag{22}
\end{equation}
Clearly $X_1^{(2)}$, $X_2^{(2)}$, $X_3^{(2)}$ satisfy the required
commutation relations. The other commutation relation gives us the partial
differential equations:

\begin{eqnarray*}
x_2\frac{\partial P_1}{\partial x_1}-x_1\frac{\partial P_1}{\partial x_2}%
-P_2 &=&x_2, \\
x_2\frac{\partial P_2}{\partial x_1}-x_1\frac{\partial P_2}{\partial x_2}%
+P_1 &=&-x_1, \\
x_2\frac{\partial P_3}{\partial x_1}-x_2\frac{\partial P_3}{\partial x_2}
&=&1,
\end{eqnarray*}
for the unknown functions $P_1,P_2,P_3$. The solution for these equations
defines the required vector to get the new Poisson structure to construct
the decomposable tensor. We can re-write the equations as:

\begin{eqnarray}
(x_2\frac \partial {\partial x_1}-x_1\frac \partial {\partial x_2})^2P_1+P_1
&=&0,  \tag{23a} \\
(x_2\frac \partial {\partial x_1}-x_1\frac \partial {\partial x_2})P_1-x_2,
&=&P_2  \tag{23b} \\
(x_2\frac \partial {\partial x_1}-x_1\frac \partial {\partial x_2})P_3 &=&1.
\tag{23c}
\end{eqnarray}
So, we just need to solve the equations (21a, 21c). We can solve this system
with the help of the transformation to polar coordinates: $x_1=r\cos \theta $%
, $x_2=r\sin \theta $, to get the equations:

\begin{eqnarray*}
\frac{\partial ^2}{\partial \theta ^2}p_1(r,\theta )+p_1(r,\theta ) &=&0, \\
\frac \partial {\partial \theta }p_1(r,\theta )-r\sin \theta ,
&=&p_2(r,\theta ) \\
\frac \partial {\partial \theta }p_3(r,\theta ) &=&1.
\end{eqnarray*}
With $P_j(x_1(r,\theta ),x_2(r,\theta ))=p_j(r,\theta )$. Hence, a solution
is:

\begin{eqnarray*}
P_1 &=&A(x_1^2+x_2^2)\sin (\arctan (\frac{x_2}{x_1})+B(x_1^2+x_2^2)), \\
P_2 &=&A(x_1^2+x_2^2)\cos (\arctan (\frac{x_2}{x_1})+B(x_1^2+x_2^2))-x_2, \\
P_3 &=&\arctan (\frac{x_2}{x_1})+C(x_1^2+x_2^2).
\end{eqnarray*}
This is enough to construct the second Poisson tensor which makes the
extension of $J^{(2)}$ and the particular $QBH_{(2)}^{*}$, like in the first
example, by just choosing a generator $H$ as solution of its corresponding
partial differential equation. The calculation of the extended Poisson
tensor gives us:

\begin{eqnarray*}
\wp ^{(1)} &=&(x_1\frac \partial {\partial x_2}-x_2\frac \partial {\partial
x_1})\wedge \frac \partial {\partial x_3}+ \\
&&([x_1\frac{\partial H^{(2)}}{\partial x_2}-x_2\frac{\partial H^{(2)}}{%
\partial x_1}]\frac \partial {\partial x_3}-\frac{\partial H^{(2)}}{\partial
x_3}[x_1\frac \partial {\partial x_2}-x_2\frac \partial {\partial x_1}%
])\wedge \\
&&(A(x_1^2+x_2^2)[\sin (\arctan (\frac{x_2}{x_1}))\frac \partial {\partial
x_1}+\cos (\arctan (\frac{x_2}{x_1}))\frac \partial {\partial x_2}]+ \\
&&+\arctan (\frac{x_2}{x_1})\frac \partial {\partial
x_3}+C(x_1^2+x_2^2)\frac \partial {\partial x_3}).
\end{eqnarray*}
In this case the Hamiltonian of the first Poisson structure must be a
solution of the differential equation:

\[
x_1\frac{\partial ^2H^{(2)}}{\partial x_2\partial x_3}-x_2\frac{\partial
^2H^{(2)}}{\partial x_1\partial x_3}=0 
\]

3.- Consider now the case of linear Poisson tensors: $J=X_A\wedge X_a$,
which define semi-direct extensions of Abelian Lie algebras\cite{cari} (the
two preceding examples are particular cases of this):

\begin{eqnarray*}
X_1 &=&X_A=\sum_{i,\text{ }j=1}^{n-1}A_i^jx_j\frac \partial {\partial x_i},
\\
X_2 &=&X_a=\frac \partial {\partial x_n}, \\
X_3 &=&\sum_{j=1}^nP_j(x_1,...,x_{n-1})\frac \partial {\partial x_j},
\end{eqnarray*}
when they act on functions belonging to $C^\infty (M,\Re )$; which is the
case important for us now. It is clear that we will take the tensor $J=$ $%
X_A\wedge X_a$ as our initial tensor. We can see that our vector fields
satisfy the requiered commutation relations if the following set of first
order partial differential equations (obtained after a straightforward
calculation which we can omit here) for the functions $P_j(x_1,...,x_{n-1})$
is solvable:

\begin{eqnarray*}
\sum_{k=1}^{n-1}A_k^jx_j\frac{\partial P_i}{\partial x_k} &=&A_i^j(P_j-x_j),%
\forall (i,\text{ }j):i,\text{ }j\in \{1,...,n-1\}, \\
(\sum_{i,\text{ }j}^{n-1}A_i^jx_j\frac \partial {\partial x_i})P_n &=&1.
\end{eqnarray*}
>From a theoretical point of view, an analytic solution for this set of
differential equations exists by the well-known Cauchy-Kovalevskaya theorem%
\cite{hil}. Hence, there is always (at least locally) an extension to the
Poisson tensor which defines semi-direct extensions of abelian Lie algebras
on the basis of the method which we propose (of course, for the use of the
Cauchy-Kovlevskaya theorem we must change $C^\infty (M,\Re )$ by $C^\omega
(M,\Re )$, the Banach space of analytic functions). A classification of
these algebras in dimension 1, 2, 3, 4 is given by Cari\~{n}ena et al\cite
{cari}. and for all the Poisson tensors which they consider is valid the
methodology which we propose, with just one remark: the Poisson tensor must
be of the form $X_1\wedge X_2$. Our method of extension is not the same as
the one of these authors, although we use a common property: in the second
Poisson tensor $X_H\wedge X_3$, we take $X_H$ as a derivation of the first
Poisson tensor. But this is all, because they want to get all the derivation
algebra, whereas we fix a derivation; a Hamiltonian vector field as
derivation; and construct the remaining piece: the vector field $X_3$. So,
if we know the algebra of derivations, our method can be applied to each
element of this algebra to get a different extension for each one of its
elements.

4.- Let us show that $\jmath (0,-1,-1)\simeq so(3)$. The commutation rules
for $so(3)$ are well-known to be (we use the circumflex accent to avoid any
confusion with the manifold coordinates):

\[
\lbrack \hat{x}_1,\text{ }\hat{x}_2]=\hat{x}_3\text{, }[\hat{x}_2,\text{ }%
\hat{x}_3]=\hat{x}_1\text{, }[\hat{x}_3,\text{ }\hat{x}_1]=\hat{x}_2 
\]
while those rules for the algebra $\jmath (0,-1,-1)$ are:

\[
\lbrack X_1,\text{ }X_2]=-X_H\text{, }[X_H,\text{ }X_1]=-X_2\text{, }[X_2,%
\text{ }X_H]=-X_1 
\]
the isomorphism is, clearly, the linear map: $X_H=-\hat{x}_3$, $X_2=\hat{x}%
_2 $, $X_1=\hat{x}_1$. So, any representation of $so(3)$ can be used to get
Jacobi structures, for example, the usual one of the form:

\[
X_H=x_2\frac \partial {\partial x_1}-x_1\frac \partial {\partial x_2}\text{,
\ }X_2=x_3\frac \partial {\partial x_1}-x_1\frac \partial {\partial x_3}%
\text{,\ }X_3=x_2\frac \partial {\partial x_3}-x_3\frac \partial {\partial
x_2} 
\]
the Jacobi structure for this case is $<X_1\wedge X_2,$ $X_H>$

COMMENT.-Well, in fact every 3-dimensional Lie algebra can be classified by
means of linear transformations in the manifold coordinates:

\[
X_i=\sum_ja_{ij}\hat{X}_j 
\]
where we suppose that the coefficients are just constants (hence we left out
non-linear transformations of the coordinates by general diffeomorphisms).
The conditions which we must impose on the linear transformation are: (1).-
it is an isomorphism, we mean, a bijective homomorphism of Lie algebras,
(2).- the Jacobi identity. This has been done, for example, by Bryant\cite
{brian} [p.37-40] and we refer there for further details on the reduction.

\section{Conclusions.}

We have shown how to construct extensions of Poisson tensors and
singular-quasi-bi-Hamiltonian systems on this basis. The procedure seems
much more comprehensive than the method due to Brouzet et al. although it is
valid only for those Poisson tensors of the form: $X_1\wedge X_2$ and its
contractions over sets of integrable (exact) 1-forms. In principle it is
possible to construct as many quasi-bi-hamiltonian systems as smooth scalar
generators are imaginable, with just one constraint; let us show why. A
solution for the partial differential equation $X_1X_2(dH)=0$ can be taken
as $H=I_1(\xi _1)+I_2(\xi _2)$ where each $\xi _i$ is an invariant function
of each vector field $X_i$. Hence, by the commutativity of the fields this
is a solution: $X_1X_2(I_1)+X_1X_2(I_2)=X_2X_1(I_1)+X_1X_2(I_2)=0$, because $%
X_1(I_1)=0$, $X_2(I_2)=0$, and because the functions $I_1$, $I_2$ are
arbitrary, our assertion is justified. The constraint is, clearly, to move
only along the characteristic paths defined by the invariants.

We can see one case in which the Hamiltonian will be totally arbitrary: if
the vector fields $X_1$, $X_2$ commute and anti-commute the first
Hamiltonian is arbitrary, there is no restriction on its form. However we
can see that the conditions: $X_1X_2=X_2X_1$ $X_1X_2=-X_2X_1$ are fulfilled
if we suppose that $X_1X_2=0$ an operator identity. But this is our
condition to get the first Hamiltonian, hence nothing new arises.

As a second point, we see that the method of extension is different from
that due to Cari\~{n}ena et al.\cite{cari}, because, as explained in the
example 3, the method can be applied to each element of the derivation
algebra to get different extensions of the same Poisson tensor.

\textbf{Acknowledgments.}

This work has been supported by the finnnacial aid of Centro de Estudios
Multidisciplinarios UAZ, and of the UAEM research project 1336/98. RAF and
MAG are grateful to Ana Maria Damore for her patient to read the manuscript
and to Rogelio C\'{a}rdenas Hern\'{a}ndez by his interest in the development
of the work. MAG acknowledges to Professor V. G. Makhankov for his constant
support discussions.

\end{document}